\documentclass[sigconf]{acmart}
\renewcommand\footnotetextcopyrightpermission[1]{} % removes footnote with conference information in first column
\AtBeginDocument{%
  \providecommand\BibTeX{{%
    \normalfont B\kern-0.5em{\scshape i\kern-0.25em b}\kern-0.8em\TeX}}}

\setcopyright{none} 
\begin{document}

%%
%% The "title" command has an optional parameter,
%% allowing the author to define a "short title" to be used in page headers.
\title{Learning gain differences between ChatGPT and human tutor generated algebra hints}

%%
%% The "author" command and its associated commands are used to define
%% the authors and their affiliations.
%% Of note is the shared affiliation of the first two authors, and the
%% "authornote" and "authornotemark" commands
%% used to denote shared contribution to the research.
% \authornote{Both authors contributed equally to this research.}
\orcid{0000-0002-6016-7051}
\author{Zachary A. Pardos}
\email{pardos@berkeley.edu}
\affiliation{%
  \institution{Berkeley School of Education}
  \city{University of California, Berkeley}
  % \state{CA}
   \country{USA}
  % \postcode{94720}
}

\author{Shreya Bhandari}
\email{shreya.bhandari@berkeley.edu}
\affiliation{%
  \institution{Electrical Engineering and Computer Science}
  \city{University of California, Berkeley}
  % \state{CA}
   \country{USA}
  % \postcode{94720}
}

%%
%% By default, the full list of authors will be used in the page
%% headers. Often, this list is too long, and will overlap
%% other information printed in the page headers. This command allows
%% the author to define a more concise list
%% of authors' names for this purpose.
\renewcommand{\shortauthors}{Pardos \& Bhandari}

%%
%% The abstract is a short summary of the work to be presented in the
%% article.
\begin{abstract}
Large Language Models (LLMs), such as ChatGPT, are quickly advancing AI to the frontiers of practical consumer use and leading industries to re-evaluate how they allocate resources for content production. Authoring of open educational resources and hint content within adaptive tutoring systems is labor intensive. Should LLMs like ChatGPT produce educational content on par with human-authored content, the implications would be significant for further scaling of computer tutoring system approaches. In this paper, we conduct the first learning gain evaluation of ChatGPT by comparing the efficacy of its hints with hints authored by human tutors with 77 participants across two algebra topic areas, Elementary Algebra and Intermediate Algebra. We find that 70\% of hints produced by ChatGPT passed our manual quality checks and that both human and ChatGPT conditions produced positive learning gains. However, gains were only statistically significant for human tutor created hints. Learning gains from human-created hints were substantially and statistically significantly higher than ChatGPT hints in both topic areas, though ChatGPT participants in the Intermediate Algebra experiment were near ceiling and not even with the control at pre-test. We discuss the limitations of our study and suggest several future directions for the field. Problem and hint content used in the experiment is provided for replicability.
\end{abstract}

%%
%% The code below is generated by the tool at http://dl.acm.org/ccs.cfm.
%% Please copy and paste the code instead of the example below.
%%

%%
%% Keywords. The author(s) should pick words that accurately describe
%% the work being presented. Separate the keywords with commas.

%% A "teaser" image appears between the author and affiliation
%% information and the body of the document, and typically spans the
%% page.
% \begin{teaserfigure}
%   \includegraphics[width=\textwidth]{sampleteaser}
%   \caption{Seattle Mariners at Spring Training, 2010.}
%   \Description{Enjoying the baseball game from the third-base
%   seats. Ichiro Suzuki preparing to bat.}
%   \label{fig:teaser}
% \end{teaserfigure}

%%
%% This command processes the author and affiliation and title
%% information and builds the first part of the formatted document.
\maketitle
\pagestyle{plain}
\section{Introduction}
ChatGPT\footnote{\url{https://chat.openai.com/chat}} has sparked debate over the range of content it and other Large Language Models (LLMs) like it can competently produce \cite{gozalo2023chatgpt,thorp2023chatgpt}. Popular discussion of ChatGPT in the educational community has centered around the concern that it could pose an existential threat to traditional assessments, should the quality of its answers be sufficient enough to score highly on many assignments \cite{rudolph6chatgpt}. To the extent that this is true, we hypothesize that ChatGPT generated answers to problems, with work shown, could also be effective for learning, serving as "Worked Solution" hints in computer tutoring systems. This style of solution hinting in algebra has been shown to lead to learning gains among secondary students\cite{kim2009tutored} and Mechanical Turk workers \cite{o2019automatic} in algebra tutoring systems. 

We investigate if ChatGPT generated hints can be beneficial to algebra learning by conducting an online experiment with 77 participants from Mechanical Turk. In our 2 x 2 design, participants are randomly assigned to the manual hint or ChatGPT generated hint condition and randomly assigned to one of two algebra tutoring lessons with questions adopted from OpenStax Elementary Algebra and Intermediate Algebra textbooks\footnote{\url{https://openstax.org/subjects/math}}. We use a soon-to-be released tutoring system, called Open Adaptive Tutor (OATutor), and its pre-made human authored hints based on this same content as the control and replace the human hints with ChatGPT produced hints to serve as the experiment condition to answer the following research questions:
\begin{itemize}
    \item \textbf{RQ1:} How often does ChatGPT produce low quality hints? 
    \item \textbf{RQ2:} Do ChatGPT hints produce learning gains?
    \item \textbf{RQ3:} How do ChatGPT hints compare to human tutor hints in learning gain?
\end{itemize}

While tutor authoring tools have improved the efficiency with which humans can transcribe tutoring content \cite{aleven2006cognitive,razzaq2009assistment}, the creative process of generating content is still labor intensive. Should ChatGPT, or other LLM-based hints be effective automatic hint generators, it would open the door to previously unrealized scaling of computer tutoring approaches in a multitude of domains and learning contexts. We make both the tutor code and all content involved in the experiment available for full reproducibility\footnote{\url{https://cahlr.github.io/OATutor-LLM-Studies}} of what we believe to be the first experiment evaluating LLM-based hints for learning gains. 
\section{Related Work}
Nascent works have conducted offline evaluation of using GPT-3 \cite{brown2020language}, the predecessor to the LLM ChatGPT is based on, in computer science education to automatically generate code explanations \cite{macneil2022generating,macneil2022automatically}. It has also been applied to math word problems and evaluated on its ability to generate variations of a word problem. Below, we present a literature review of work using other methods to automatically generate hints and provide additional background on LLMs.
\subsection{Automatic hint generation}
Past work has grappled with the role of data in automatically generating hints, but following a common intuition that successful paths observed in the past can be synthesized to help guide future learners. This approach was applied to a logic tutor, modeling student past paths as a Markov Decision Process (MDP) \cite{barnes2008toward} and demonstrating positive learning outcomes when piloting the approach in practice \cite{stamper2013experimental}. 

Computer programming has been a particularly active domain for exploring automatic hint generation \cite{price2019comparison}. \citet{rivers2014automating} suggested an approach whereby programming solution states are mapped from a mixture of verbatim past observed states and canonicalized states, produced by removing syntactic differences among semantically similar states. \cite{piech2015autonomously} presented a data-driven problem solving policy evaluation framework with experiments run on Code.org's Hour of code data, finding that programming solution paths were better modeled with a Poisson policy than as an MDP as modeled in the logic tutor. \cite{10.1145/3231644.3231690} argued for using heuristics that mimic experts for generating hints and showed marginal improvements over purely data-driven approaches applied to the same Code.org dataset. Hint generation has also been explored for open-ended programming assignments \cite{price2016generating} and for coding style improvement \cite{roy2016scale}. 

\subsection{Large Language Models}
Highly parameterized neural networks trained on very large text corpora mark the current generation of Large Language Models (LLMs). These models also have in common the foundation model \cite{bommasani2021opportunities} architecture of the Transformer \cite{vaswani2017attention}, which in 2017 introduced the attention mechanism, applied in subsequent Natural Language Processing models to effectively infer word meaning based on sentence context. Both GPT \cite{radford2018improving} and the popular BERT \cite{devlin2018bert} and SentenceBERT \cite{reimers2019sentence} models share the Transformer as their base architecture, with GPT utilizing decoding components (i.e., generating oriented) and BERT utilizing encoding components (i.e., embedding oriented) of the architecture. The breakthrough in ChatGPT's usability comes from a combination of its intuitive and currently freely accessible interface and its use of a GPT model that has undergone several stages of evolution, with the most recent stage making use of human raters to better align the model's generated text to responses rated as desirable \cite{radford2018improving,radford2019language,brown2020language,ouyang2022training}. 
\section{Methods}

\subsection{Subject lesson selection} 
High school Algebra was selected as it is the most studied subject for tutoring and thus the subject with the most existing baselines and content to compare to. It is also the subject for which pre-authored questions and hints were available under a CC B-Y license from the OATutor system. To decide which objectives would be utilized in the study, each OATutor question was examined in terms of its associated objective. Within Algebra, we decided to choose one lesson from the Elementary Algebra and one from Intermediate Algebra textbooks. Each textbook is comprised of chapters, which contain learning objectives, and sets of problems belonging to those learning objectives which we are calling lessons.

Consistent with past algebra experiments, we set a requirement to have a three item pre-test and repeated post-test, and a five item acquisition phase. This meant a minimum of eight problems had to be associated with a learning objective for it to qualify for inclusion in the study. Additionally, none of the problems could depend on any images or figures, since a current limitation of ChatGPT and most other LLMs is that they only support text as input and output. Skipping chapter one in each book, because it covers prerequisite content, we advanced through each chapter and learning objective in order until we found a learning objective that satisfied the criteria. This resulted in the selection of \textit{Solve Equations Using the Subtraction and Addition Properties of Equality} as the learning objective from Chapter 2.1 of Elementary Algebra and \textit{Solve linear equations using a general strategy} from Chapter 2.1 of Intermediate Algebra. 
\subsection{ChatGPT Hint Generation}
\subsubsection{Model}
ChatGPT is a chat interface to a machine learning model based on the Generative Pre-trained Transformer (GPT) architecture. Fundamentally, ChatGPT takes as input a block of text produced by the user (e.g., "What were the best movies of the 1980s?") and returns a block of text in response. In this scenario, the input text, referred to as a "prompt," is used to inference the model which has already been pre-trained on a massive corpora of text. Prior language model approaches to this prompt/response scenario treated the response as a text completion of the prompt. However, the ways in which users interact with language models for a desired response differ from the ways those prompt/response pairs tend to manifest in the training corpora. For example, text in the corpora is likely to contain a list of movies (i.e., the desired response) following the text, "The best movies of the 1980s were...," (i.e., the prompt). However, users interacting with LLMs do not tend to use that style of text completion prompt, but instead prefer to query with a prompt posed as a question or an instruction (e.g., "Please tell me the best movies of the 1980s"). This observation of the misalignment between the training data and user prompts led to a process of alignment using human generated responses to prompts and ratings of GPT responses. This alignment, using reinforcement learning from human feedback (RLHF) \cite{christiano2017deep}, produced a model called InstructGPT (or GPT 3.5) \cite{ouyang2022training}, the basis for ChatGPT. In our study, the December 15th, 2022 version of the model is used to produce problem hints for our experiment condition.

\subsubsection{Prompt engineering}
\begin{figure}
\caption{ChatGPT hint example}
  \label{fig:chatgpthint}
    \includegraphics[width=\linewidth]{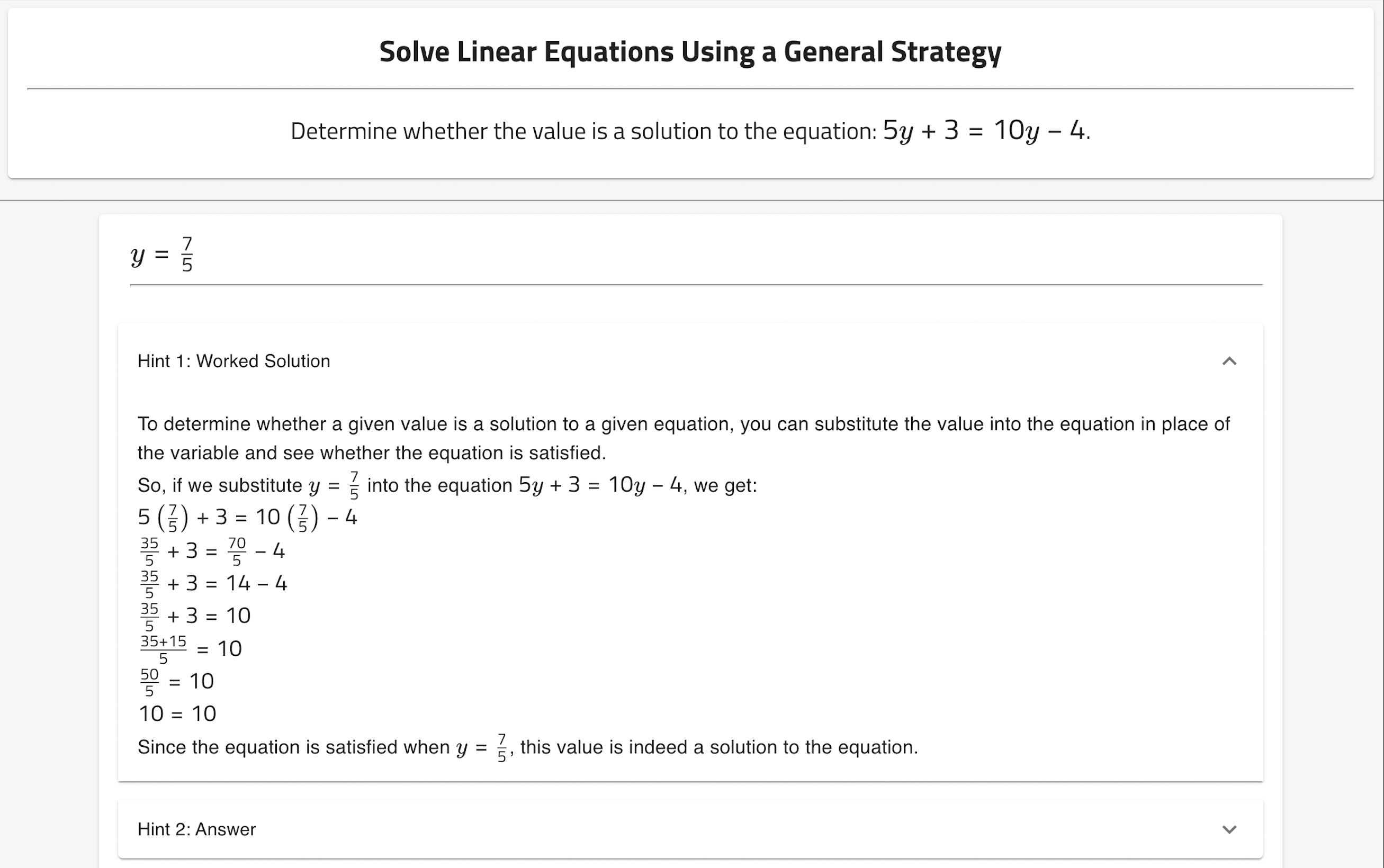}
\end{figure}

For every problem in the two selected lessons, we posed the question of the problem to ChatGPT directly and recorded its response to potentially serve as a hint. A problem and example ChatGPT hint for the problem is shown in Figure \ref{fig:chatgpthint}. The prompt was a concatenation of the text components of a problem defined by OATutor (i.e., <problem header> <problem body> <step header> <step body>). When providing the prompt for a new question, a "New chat" was created to clear the history and prevent the model from potentially using information from the previous prompt. We explored following up with a second prompt of "Please explain" to see if a different quality of response would be given. This was considered as a potential third experimental condition, but since the response was so similar to the original response, we did not pursue it further.  

\subsubsection{Quality checks}
Large Language Models are known to sometimes produce plausible or confident statements that are factually incorrect \cite{shuster2021retrieval,maynez-etal-2020-faithfulness}. To prevent incorrect or potentially inappropriate hint content from making its way to study participants, we conducted a quality check of all ChatGPT generated hints. This consisted of a three point check to ensure that 1) the correct answer was given in the worked solution 2) the work shown was correct and 3) that no inappropriate language was used. A hint was considered fully correct if it met these three criteria. If a hint failed to meet even one of these criteria, the question it was associated with was disqualified, resulting in a decrease in the pool of questions that could be utilized for the experiment. After this process, if the number of questions that were not disqualified was greater than or equal to 8 questions, then the associated objective would be utilized for the study. However, if less than 8 questions resulted, then a new objective (associated with the same OpenStax book as the original objective) was chosen and the whole process detailed in this study was repeated with the new objective. The time taken to conduct this quality check and rejection statistics were recorded to later consider as part of the cost of using ChatGPT for hint generation and are reported in Table \ref{tab:quality}.

\begin{table*}[hbt!]
\caption{ChatGPT quality check results}
  \label{tab:quality}
\begin{tabular}{cccccccc} 
    \toprule
 Textbook Level & N & Quality Check Time & \# Disqualified & \# Inappropriate & \# Incorrect Work & \# Incorrect Answer\\ 
    \midrule
 Elementary & 24 & 11 min, 0 sec & 5 & 0 & 5 & 5\\ 
 Intermediate & 16 & 7 min, 42 sec & 7 & 0 & 7 & 7\\
     \bottomrule

\end{tabular}
\end{table*}

\begin{table*}[hbt!]
  \caption{Study learning gain results}
  \label{tab:results}
\begin{tabular}{cccccccc} 
    \toprule
 Textbook Level  & Condition & N & Avg. Time & Hints Requested & Learning Gain & Avg. Pre-test & Avg. Post-test\\ 
    \midrule
 Elementary & Control & 19 & 08:16 & 132 & 24.63\% & 59.68\% & 84.32\%\\ 
 Elementary & Experiment & 21 & 09:01 & 30 & 11.14\% & 74.67\% & 85.81\%\\
 Intermediate & Control & 17 & 12:53 & 150 & 23.65\% & 50.94\% & 74.59\%\\
 Intermediate & Experiment & 20 & 11:06 & 57 & 1.7\% & 80.05\% & 81.75\%\\
     \bottomrule

\end{tabular}
\end{table*}

\subsection{Manual Hint Generation}

% For constructing manual hints, a Google Spreadsheet was created for each book (Elementary Algebra, Intermediate Algebra, College Algebra, and Statistics) with each sheet corresponding to a different lesson. Each sheet had the following columns: Problem Name, Row Type, Title, Body Text, Answer, answerType, HintID, Dependency, mcChoices, Images, OER src, openstax KC, KC, and Taxonomy. 

% The “problem name” column simply contained the lesson name and problem number. In the “row type” column, there were 4 options as inputs. The different row types and their definitions are provided below:
% problem: Rowtype utilized to start a new problem (could have multiple questions under the same problem)
% step: Rowtype utilized to start a new question
% hint: Rowtype utilized to provide the OAT user with a hint (no answer is allowed)
% scaffold: Rowtype utilized to provide the OAT user with a hint in the form of a question (answer is necessary)
% Based on the row type, the subsequent columns could be filled.

\begin{figure}
\caption{Manually generated hint example}
  \label{fig:manualhint}
    \includegraphics[width=\linewidth]{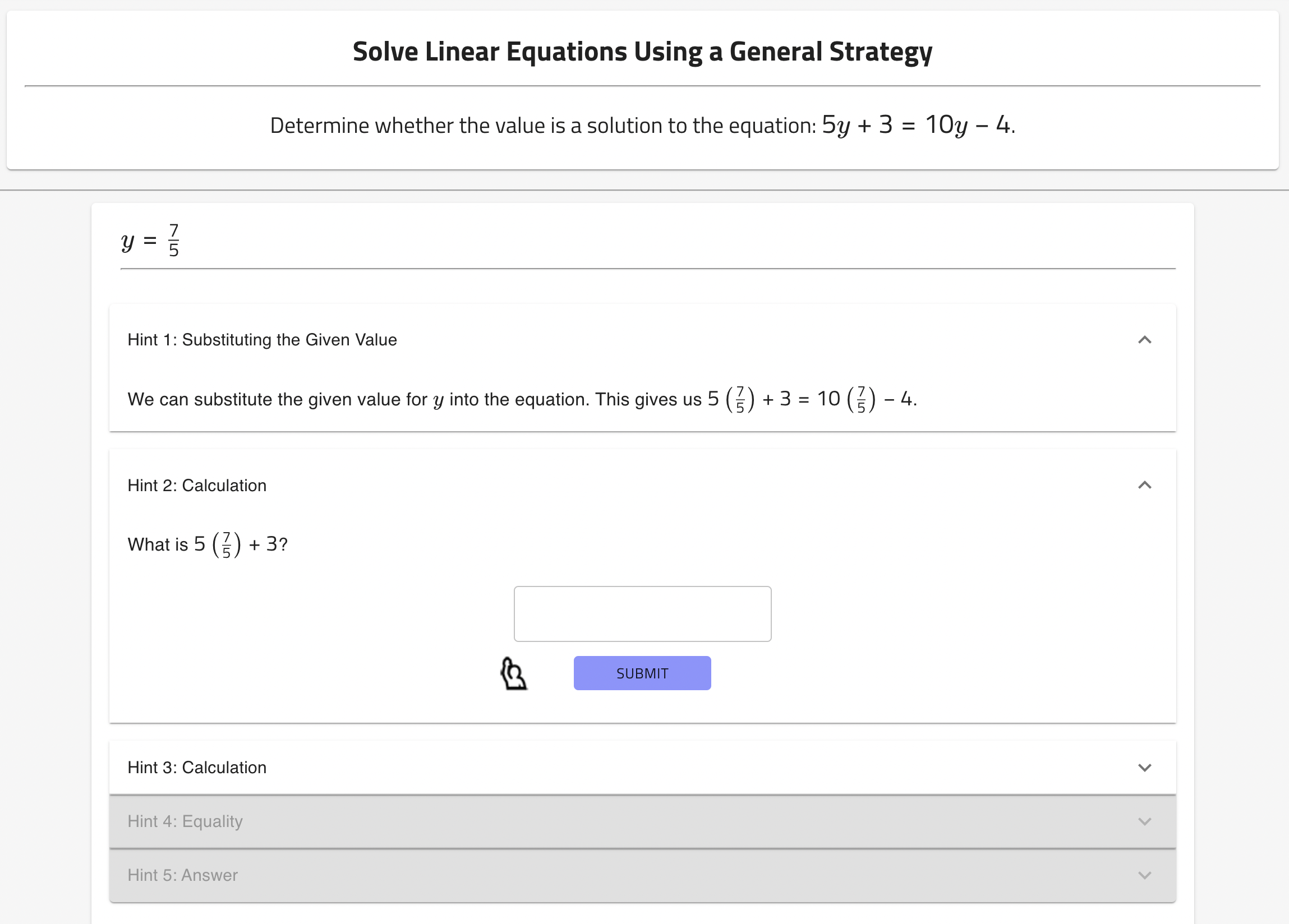}
\end{figure}

We utilized the already created human tutor hints in the OATutor system. Those hints were produced by undergraduate students with prior math tutoring experience. The system allowed tutor authors to enter any combination of text hints or hints in the form of a question that breaks the problem down into a small subtask, called a scaffold. There was no limit to the number of hints/scaffolds a particular step would have. The authored content was quality checked by editors on the OATutor content team, though the time taken for this quality check was not reported. An example manually generated set of hints is shown in Figure \ref{fig:manualhint} for the same problem as the ChatGPT hint example.

\section{Experiment Setup}
\subsection{Experimental design}
Each of the four experimental conditions consisted of a three item pre-test, followed by a five item acquisition phase, and finishing with a three item post-test consisting of the exact same items as the pre-test. The participant was first shown a consent screen, followed by being randomly assigned to either the control (i.e., manual hints) or experiment (i.e., ChatGPT hints) and either the Elementary Algebra or Intermediate Algebra lesson. The experiment ended by showing the participant a survey code representing their anonymized user ID in the OATutor system and then a thank you screen. The OATutor system handled logging of the condition, lesson name, anonymized user ID, problem name, correctness of response, hint request actions, and timestamp, which we directed to our own Firebase account for later download and analysis.

\subsection{Participants}
Amazon's Mechanical Turk marketplace was utilized to recruit participants. In Mechanical Turk, we limited the participants only to those who had at least a high school degree and had the "master" designation, meaning they had demonstrated a successful record of task completion on the platform. The high school requirement was placed for two reasons. Firstly, individuals with a high school degree would likely have knowledge of the prerequisite topics to elementary algebra and intermediate algebra. This means that after learning the content through the hints/feedback during the condition phase, learning gain may be possible from the pre-test to the post-test phases. Additionally, since Mechanical Turkers may not have been exposed to math problem solving recently, there is a better chance of seeing improvement in their scores after relearning the concepts through the hints/feedback. The compensation given to Mechanical Turkers was 8 dollars for an expected session of 10-20 minutes. The target number of participants was 20 participants per lesson and condition pairing, resulting in an overall target of 80 participants.

\section{Results}
The recruitment resulted in 77 participants who completed their lesson, with only three MTurk participants needing to be thrown out. Learning gain results from the four experiment conditions are shown in Table \ref{tab:results}, as well as statistics on average time spent in the lesson per participant, number of total hints requested across all participants in the condition, and average pre and post-test scores. The learning gain is calculated as the average post-test average score subtracted by pre-test average score for each participant. The Mann Whitney U test of statistical significance was chosen for all comparisons since the null hypothesis of normality for learning gains, pre-test, and post-test scores was rejected using a Shapiro-Wilk test. For both the Elementary Algebra and Intermediate Algebra lessons, learning gains were higher in the control condition and statistically significantly separable from the experiment (both with p = 0.038). Participants in the control and experiment conditions were even at pre-test in Elementary Algebra (p = 0.1598) but not in Intermediate Algebra (p = 0.0029), where participants in the control scored a 50.94\% and 80.05\% in the experiment. Finally, all experiments showed positive learning gains; however, they were statistically significant differences only between pre and post-test scores of the control condition (p = 0.0219 for Elementary Algebra and p = 0.0213 for Intermediate Algebra), but not the experiment condition (p = 0.1427 for Elementary Algebra and p = 0.7912 for Intermediate Algebra). %INSERT SENTENCE ABOUT STAT SIG OF PRE vs POST TEST (i.e., Was there learning RQ2)

\balance
\section{Conclusions and Discussion}
Results of our study producing algebra hints using ChatGPT showed a 30\% rejection rate of produced hints based on quality (RQ1), suggesting that the technology still requires human supervision in its current form. All of the rejected hints were due to containing the wrong answer or wrong solution steps. None of the hints contained inappropriate language, poor spelling, or grammatical errors. Our experiments, comparing learning gain differences between ChatGPT generated hints and manually generated hints, showed that all experiments produced learning gains; however, they were only statistically significant among the manual hint conditions (RQ2). Manual hints produced higher learning gains than ChatGPT hints in both lessons and these differences were statistically significantly separable (RQ3). However, participants in the experiment condition for Intermediate Algebra were near ceiling at pre-test (80.05\% avg.) and statistically significantly separable from the pre-test scores of the participants in the control (50.94\% avg.). A similar amount of time was spent between both control and experiment in the two lessons, indicating that while the number of hints requested in the control was much greater, because manual hints were unbounded in the number that could be authored, there was not a time savings by seeing fewer hints in the experiment conditions. 

Worth considering, is if this result is indicative of the quality difference between machine and human created hints or if it reflects the difference in efficacy between worked solutions and the use of hints and scaffolds. Future work could isolate this by comparing to manually generated worked solutions or having ChatGPT generate scaffolding and several hints through prompt engineering. Additionally, conflated in the learning gains is the effect of immediate feedback (i.e., being told if an answer is correct or incorrect). Further isolation of the effect of the hints could be achieved by adding an immediate feedback-only condition, whereby students are not shown any hints but are told the correctness of their response.

Mechanical Turkers proved suitable for this experiment, with a high completion rate (77 out of 80) and an overall average post-test gain in all experiments. Due to the high variability of background knowledge exhibited in the pre-test scores for the same subject (50.94\% control v.s. 80.05\% in experiment for Intermediate Algebra), a higher N size may be called for. Given Turkers pre-test scores were all above 50\%, experiments involving more advanced material may also be appropriate in order to mitigate against ceiling effects. Ideal experimental circumstances would involve students as participants, right as the topics are being introduced in their curriculum. However, this coordination is very difficult to achieve in secondary schools, particularly at scale.
    
Future work could incorporate nascent advancements in LLMs \cite{li2023blip} that may allow for the use of images in problems and potential to reduce the hint generation rejection rate through self-consistency \cite{wang2022self}. Future work may also explore personalization of ChatGPT generated hints and expansion to more advanced topic areas in mathematics as well as outside of STEM.

\begin{acks}
This study was approved by the UC Berkeley Committee for the Protection of Human Subjects under IRB Protocol 2022-12-15943.
\end{acks}

%%
%% The next two lines define the bibliography style to be used, and
%% the bibliography file.
%%% -*-BibTeX-*-
%%% Do NOT edit. File created by BibTeX with style
%%% ACM-Reference-Format-Journals [18-Jan-2012].

\end{document}